\documentclass[aps,prb,superscriptaddress, twocolumn]{revtex4-1}
\usepackage{braket}
\usepackage{graphicx}
\usepackage{subfig}
\usepackage{amsmath}
\usepackage[justification=centerlast]{caption}

\begin{document}
\graphicspath{}

\title{Fractional Chern Insulator phase at the transition between checkerboard and Lieb lattices}


\author{B\l a\.{z}ej Jaworowski}
\affiliation{Department of Theoretical Physics, Wroclaw University of Technology, Wybrzeze Wyspianskiego 27, 50-370 Wroclaw, Poland}
\author{Andrei Manolescu}
\affiliation{School of Science and Engineering, Reykjavik University, Menntavegur 1, IS-101 Reykjavik, Iceland}
\author{Pawe\l ~Potasz}
\affiliation{Department of Theoretical Physics, Wroclaw University of Technology, Wybrzeze Wyspianskiego 27, 50-370 Wroclaw, Poland}


\date{\today}

\begin{abstract}
The stability of $\nu=1/3$ Fractional Chern Insulator (FCI) phase is analysed on the example of checkerboard lattice undergoing a transition into Lieb lattice. The transition is performed by the addition of a second sublattice, whose coupling to the checkerboard sites is controlled by sublattice staggered potential. We investigate the influence of these sites on the many body energy gap between three lowest energy states and the fourth state. We consider cases with different complex phases acquired in hopping and a model with a flattened topologically nontrivial band. We find that an interaction with the additional sites either open the single-particle gap or enlarge the existing one, which translates into similar effect on the many-particle gap. Evidences of FCI phase for a region in a parameter space with larger energy gap are shown by looking at momenta of the three-fold degenerate ground state, spectral flow, and quasihole excitation spectrum.
\end{abstract}


\maketitle

Recent work on Fractional Chern Insulators (FCI) as a lattice version of Fractional Quantum Hall Effect (FQHE) \cite{tsui, laughlin} without a need of Landau levels has attracted significant attention\cite{Sun,tang,Neupert,wang,SunNature,PRX, hierarchy,RevParameswaran,RevBergholtz,RevNeupert}. Those are many-particle extension of Chern insulators\cite{Haldane} - systems which exhibit integer quantum Hall effect without magnetic field and were recently realized experimentally \cite{ChernExperiment, ChernExperiment2}. FCI are particularly interesting because they can mimic Landau level physics and may provide a more convenient way of conducting experiments on FQHE, as they can exist in higher temperature and would not need high magnetic fields\cite{tang}. FCI can also depart from Landau level physics, which happens e. g. for bands with Chern number higher than one, where new forms of FCI states can arise \cite{highchern1,highchern2,highchern3,highchern4}.

Experimental realizations of FCI phase were proposed in different systems including cold atoms\cite{opticalflux1} or molecules in optical lattices\cite{dipolarspin1, dipolarspin2}, graphene \cite{MultilayerGraphene, StrainedGraphene, FloquetGraphene}, arrays of quantum wires \cite{wires}, transition-metal oxide heterostructures\cite{NatureTransMet, dice}, or strongly-correlated electrons in layered oxides\cite{KourtisTriangular,layers1, layers2}. 

Initially, it was proposed that FCI should exist on topologically nontrivial flat band models \cite{Sun,tang,flatkagome,flatstar,flatsqoct, highchern3}. Several lattice models with quasi-flat topologically nontrivial bands have been shown numerically to exhibit FCI phase, including checkerboard \cite{Neupert,PRX, SunNature,wang,YangDisorder}, honeycomb\cite{zoology,wang}, square\cite{zoology}, triangular\cite{KourtisTriangular}, and Kagome lattices\cite{zoology}. Numerical evidence for analogs of a number of FQHE states, including Laughlin $1/m$ \cite{Neupert}, CF hierarchy \cite{hierarchy, BeyondLaughlin} and non-Abelian Moore-Read and Read-Rezayi states \cite{BernevigCounting,MooreReadWang, zoology} was found. For bands with higher Chern number, states with no direct analog in FQHE were found, some of which exhibiting non-Abelian statistics\cite{highchern1,highchern2,highchern4}. 

To prove existence of FCI in torus geometry for filling $p/q$ one should show $q$ quasi-degenerate ground states \cite{wen,HaldaneCounting, quasideg}, which flow into each other and do not intersect with higher states when one flux quantum is inserted through a handle of the torus\cite{LaughlinArgument,sflow1, quasideg}, and obey the momentum counting rules\cite{HaldaneCounting, BernevigCounting}. These rules need to be satisfied also for quasihole excitations\cite{zoology, PRX}. Alternative methods of proving FCI existence include many-body Chern number \cite{sflow1, quasideg,Neupert} and entanglement spectrum \cite{Entanglement,EntanglementExcitation, ParticleEntanglement,PRX}. 

There are several criteria which allow to find systems which can host FCI phase. First, the flatness ratio (a ratio of magnitude of band dispersion to the energy gap) needs to be low, to maximize the effect of interaction. However, this criterion has proven ambiguous, as the single-particle dispersion can stabilize the FCI phase \citep{KourtisTriangular,SingleParticleGrushin,SingleParticleChamon,SingleParticleMurthy, hierarchy}, and interactions far exceeding band gap do not always lead to destruction of FCI \cite{FarExceed}. Secondly, in the limit of long wavelength and uniform Berry curvature, the projected density operator algebra resembles the Girvin-MacDonald-Platzman algebra \cite{Girvin} for a Landau level. In consequence an energy band needs to have nearly-flat Berry curvature to host FCI phase\cite{zoology, BerryParameswaran}.  Also, a third criterion, based on Fubini-Study metric was proposed recently \cite{FubiniRoy,FubiniStudy, FubiniNeupert}. However, clear conditions for FCI existence are not perfectly understood.

In this work, we want to investigate how the stability of FCI on a given lattice is affected by introducing an interaction with extra lattice sites. We consider a checkerboard lattice which transforms into a Lieb lattice\cite{Lieb,Aldea,Weeks,Zhao,Goldman} when a second sublattice is introduced into the system, controlled by on-site staggered potential. We investigate the transition between two lattices in the context of FCI phase for spinless particles for $1/3$ filling. For finite-size systems in a torus geometry, we analyze the influence of the interaction between the two sublattices on the many-body energy gap between three lowest energy states and the fourth state. For a specific choice of parameters corresponding to an area of larger energy gap, we search for signatures of $1/3$ Laughlin-like phase. Three lowest energy states (a three-fold ground state manifold) are analyzed with respect to (i) their momenta, (ii) the energy gap to excited states for different systems sizes, (iii) spectral flow. Also, the quasihole spectrum and its momentum counting is investigated. Our results suggest the existence of FCI phase with a stability supported by the interaction with extra lattice sites. The paper is organized as follows: in Section I we describe the lattice model, Section II contains a single particle analysis, in Section III many-body effects are investigated, and in Section IV we conclude our results.

\section{Model}
\begin{figure}
\begin{minipage}{0.5\textwidth}
\subfloat[][]{\includegraphics[width=3.3in]{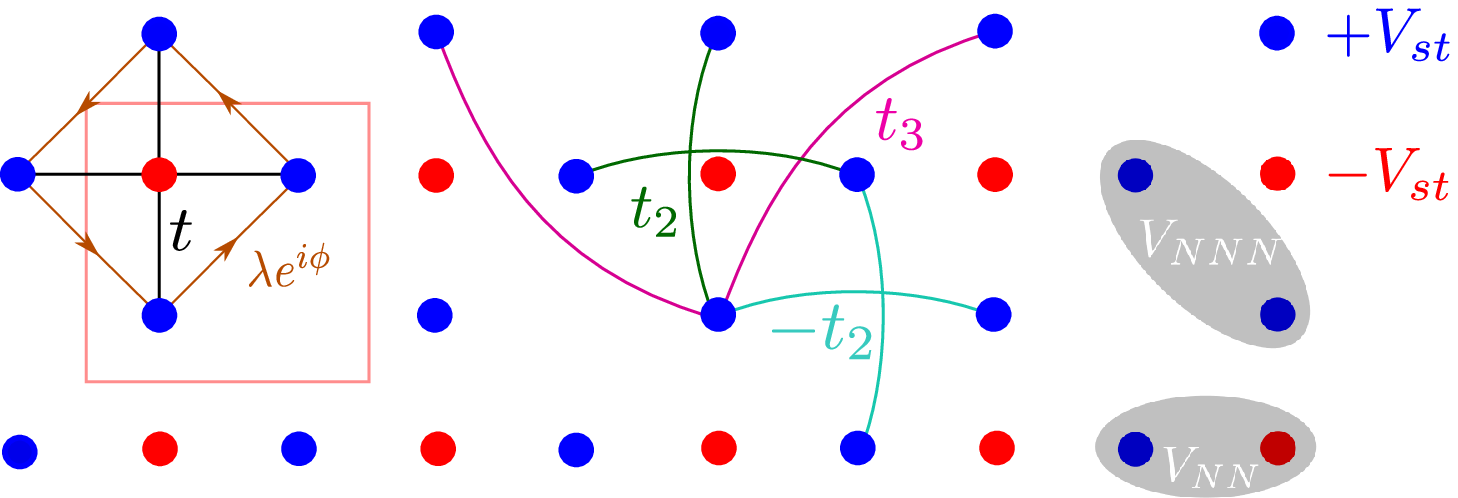}}\\
 \subfloat[][]{\includegraphics[width=3.3in]{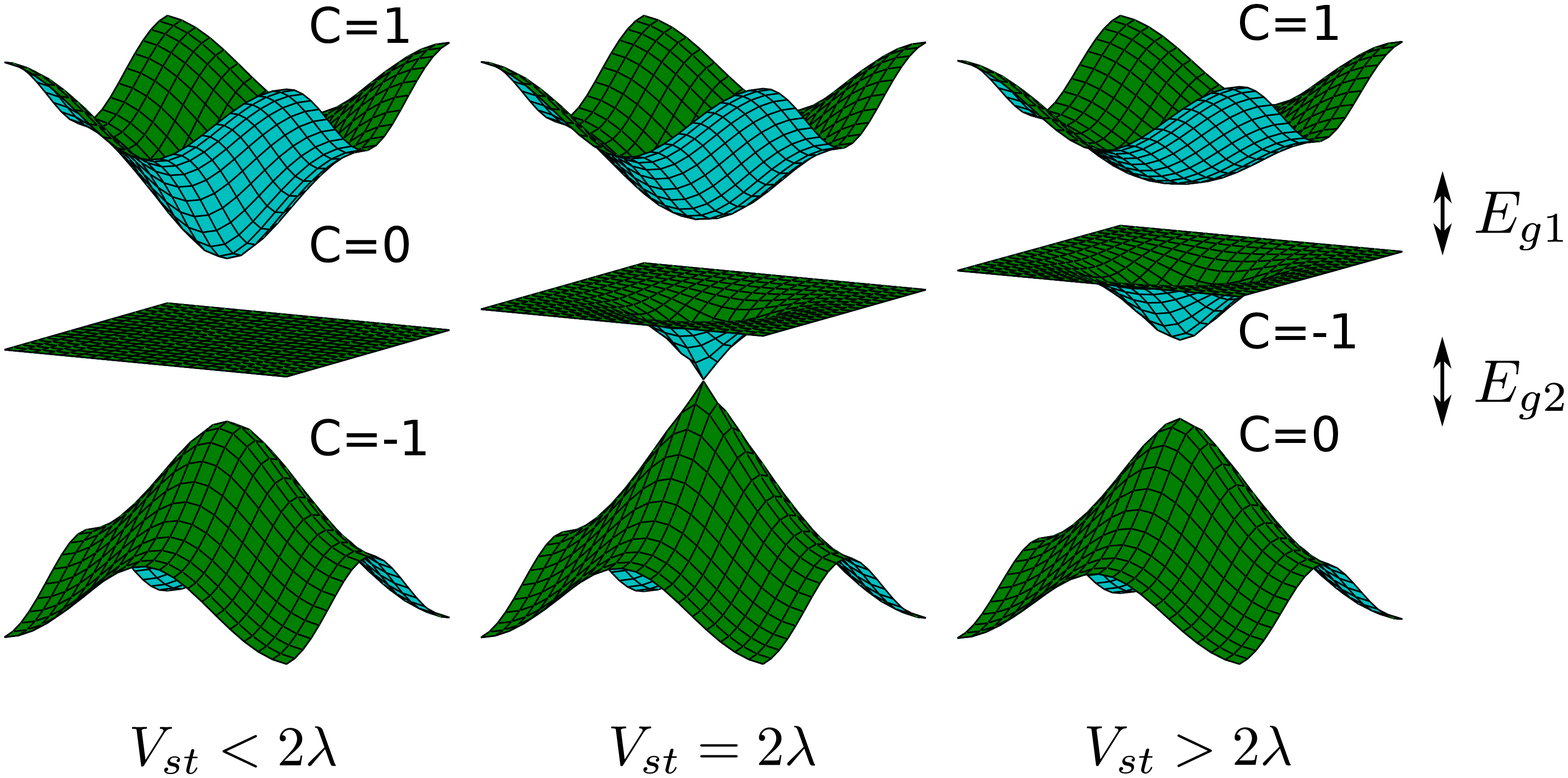}}\\
\end{minipage}
\captionsetup[subfigure]{justification=raggedright}
\caption{ (a) Structure of Lieb lattice. Red and blue atoms belong to sublattices A and B, respectively. Solid black lines denote real nearest-neighbour hoppings, arrows denote complex second-neighbour hoppings. Other solid lines denote further-neighbour hoppings used to flatten the middle band. $t_2$ hopping connects the second-nearest neighbours within B sublattice, if an A atom is between them. Otherwise, the hopping is $-t_2$. $t_3$ hoppings connect third-nearest neighbours within B sublattice. Grey ellipses denote interaction parameters. (b) Topological phase transition in Lieb lattice.}
\label{fig:transstr}
\end{figure}

A face centered 2D square lattice called a Lieb lattice is considered, shown in Fig. \ref{fig:transstr}(a). The lattice can be divided into two sublattices A and B, distinguished in Fig. \ref{fig:transstr}(a) by red and blue colors. We use  tight-binding Hamiltonian
\begin{multline}
H=t\sum_{\braket{i,j}}c^{\dagger}_i c_j+\lambda\sum_{\braket{\braket{i,j}}} e^{i\phi_{ij}}c^{\dagger}_{i} c_{j}+\\+V_{st}\sum_{i\in A}c^{\dagger}_{i} c_{i}-V_{st}\sum_{i\in B} c^{\dagger}_{i} c_{i},
\label{Ham}
\end{multline}
where in the first term $\braket{}$ denotes summation over nearest neighbors with the hopping integral $t$, the second term is a next-nearest neighbors term denoted by 
$\braket{\braket{}}$ with hopping amplitude $\lambda$ and an accumulated extra complex phase $\phi_{ij}=\pm\phi$ when going clockwise and counterclockwise, respectively, and $V_{st}$ is a staggered sublattice potential. We note that for $\phi=\pi/2$ the second term corresponds to Kane-Male spin-orbit coupling \cite{KaneMele}, and $\phi=\pi/4$ was considered for checkerboard lattice in Refs. \cite{Neupert,Sun}. In the latter case, extra hoppings were added to open the gap and flatten one of the bands; they are shown as $t_2$ and $t_3$ in Fig. \ref{fig:transstr}(b). Based on Refs\cite{Neupert,Sun}, these hoppings have values $t_2 = \frac{\lambda}{2+\sqrt{2}}$ and $t_3 = \frac{\lambda}{2+2\sqrt{2}}$.A transition between a Lieb lattice and a checkerboard lattice is driven by tuning $V_{st}$ to infinity. In this case, lattice sites represented by red color in Fig. \ref{fig:transstr}(b) are decoupled from sites represented by blue color, and systems consisting of sites of different colors can be treated independently, with blue color sites forming a checkerboard lattice. A systematic analysis of this transition will be presented in next Section. 

Many-body effects are studied using density-density interaction of the form
    \begin{equation}
    V=V_{NN}\sum_{\braket{i,j}}n_in_j+V_{NNN}\sum_{\braket{\braket{i,j}}}n_in_j,
\label{Vint}
    \end{equation}
where $n_i$ is a density operator on site $i$, and $V_{NN}$ and $V_{NNN}$ are interactions between first and second neighbors, respectively. We will focus on correlation effects within the middle band, so the Hilbert space is truncated, containing states from this band only. The lower band is considered as completely filled. Also, a flat-band approximation is
used neglecting the kinetic energies. We note that middle band states are localized mostly on one sublattice (indicated by blue color in Fig. \ref{fig:transstr}(b)) even for low $V_{st}$, as long as it is topologically nontrivial. Therefore, the leading term in Eq. (\ref{Vint}) is between second-neighbors, $V_{NNN}$. All calculations are performed for finite size $N_x\times N_y$ samples with a torus geometry, where $N_x$($N_y$) is a number of unit cells in $x$($y$) direction. We consider $1/3$ filling of the middle band which corresponds to $N=N_x\cdotp N_y/3$ particles in the system. Due to a translation symmetry and momentum conservation of two particle Coulomb scattering term, many-body eigenstates can be indexed by total momentum quantum numbers $K_x$ and $K_y$, which are the sum of the momentum quantum numbers of each of the N particles modulo $N_x$ and $N_y$, respectively. 

\section{Single particle analysis}
The unit cell of Lieb lattice consists of three sites giving three energy bands after diagonalization of Hamiltonian given by Eq. \ref{Ham}. A band structure in the simplest case when only nearest-neighbor hopping integrals $t$ are included has the lower and upper bands touching each other in the middle of energy spectrum at energy $E=0$, where the perfectly flat third energy band is present \cite{Weeks}. Two dispersive bands are almost equally localized on both sublattices, while the flat middle band is almost fully localized on a sublattice indicated by a blue color in Fig. \ref{fig:transstr}(a). We next introduce the second term from Eq. \ref{Ham} with $\phi=\pi/2$. The energy gap opens and the lower and upper bands are topologically nontrivial with Chern numbers $C=-1$ and $C=1$, respectively, and the middle flat band is topologically trivial with Chern number $C=0$, as shown in Fig. \ref{fig:transstr}(b) on the left. Following Zhao et al. \cite{Zhao}, the topology of the energy bands can be changed by introducing a staggered sublattice potential, i.e. the two last terms in Hamiltonian given by Eq. \ref{Ham}.  Increase of $V_{st}$ leads to bending of the middle band. At a critical value of $V_{st}=2\lambda$, the middle and lower bands touch, the band structure shown in the middle in Fig. \ref{fig:transstr}(b). At this point a topological phase transition occurs. For $V_{st}>2\lambda$, the lower band becomes topologically trivial with Chern number $C=0$, while the middle band becomes nontrivial with Chern number $C=-1$, what is shown in Fig. \ref{fig:transstr}(b) on the right. Similar transition occurs for $\phi=\pi/4$, but at the value $V_{st}=\sqrt{2}\lambda$ and at $V_{st}=\lambda$ when $t_2$ and $t_3$ are considered.

\begin{figure}
\begin{minipage}{0.5\textwidth}
    \subfloat[][]{\includegraphics[width=1.7in]{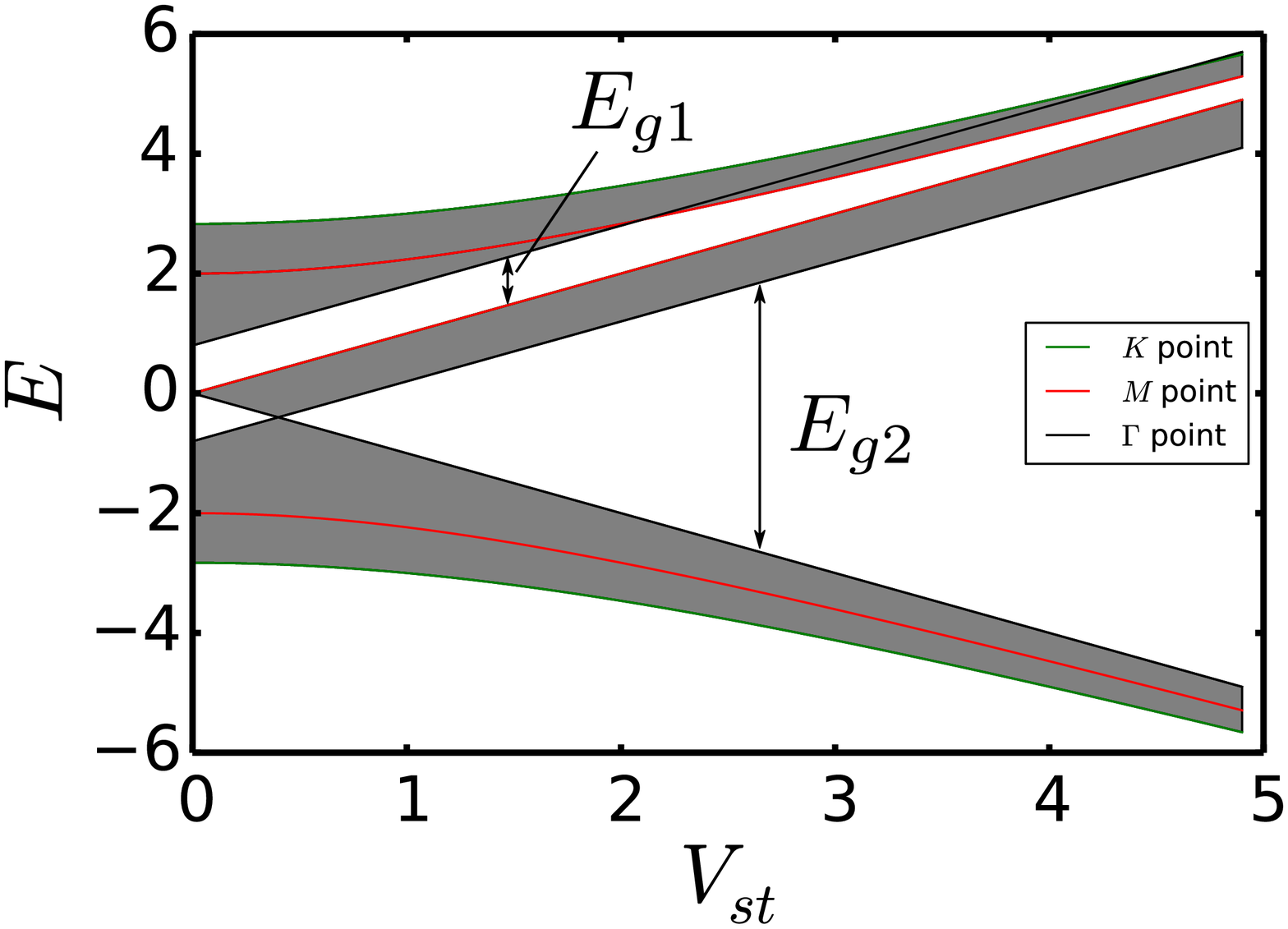}}
    \subfloat[][]{\includegraphics[width=1.7in]{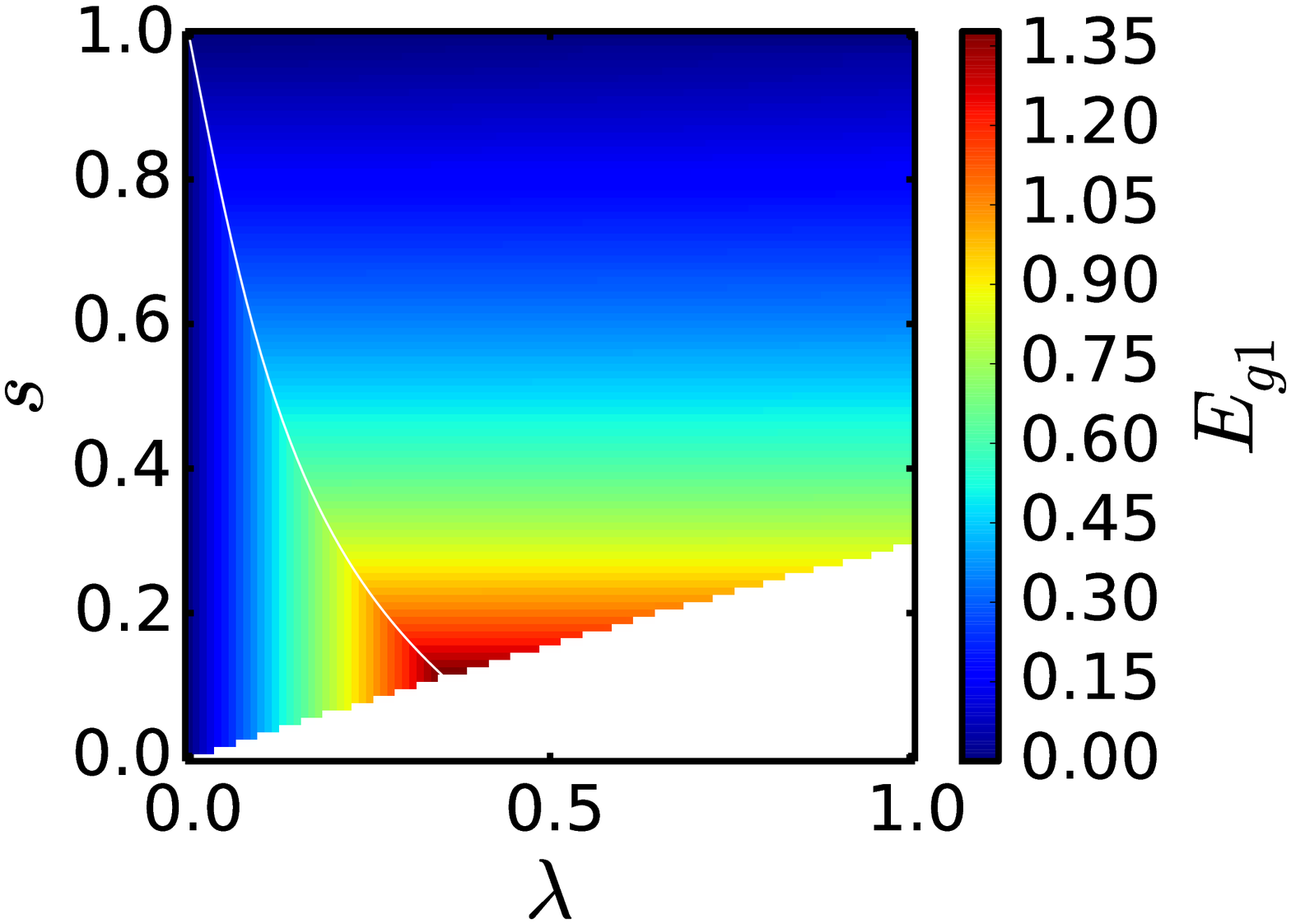}}\\
\subfloat[][]{\includegraphics[width=1.7in]{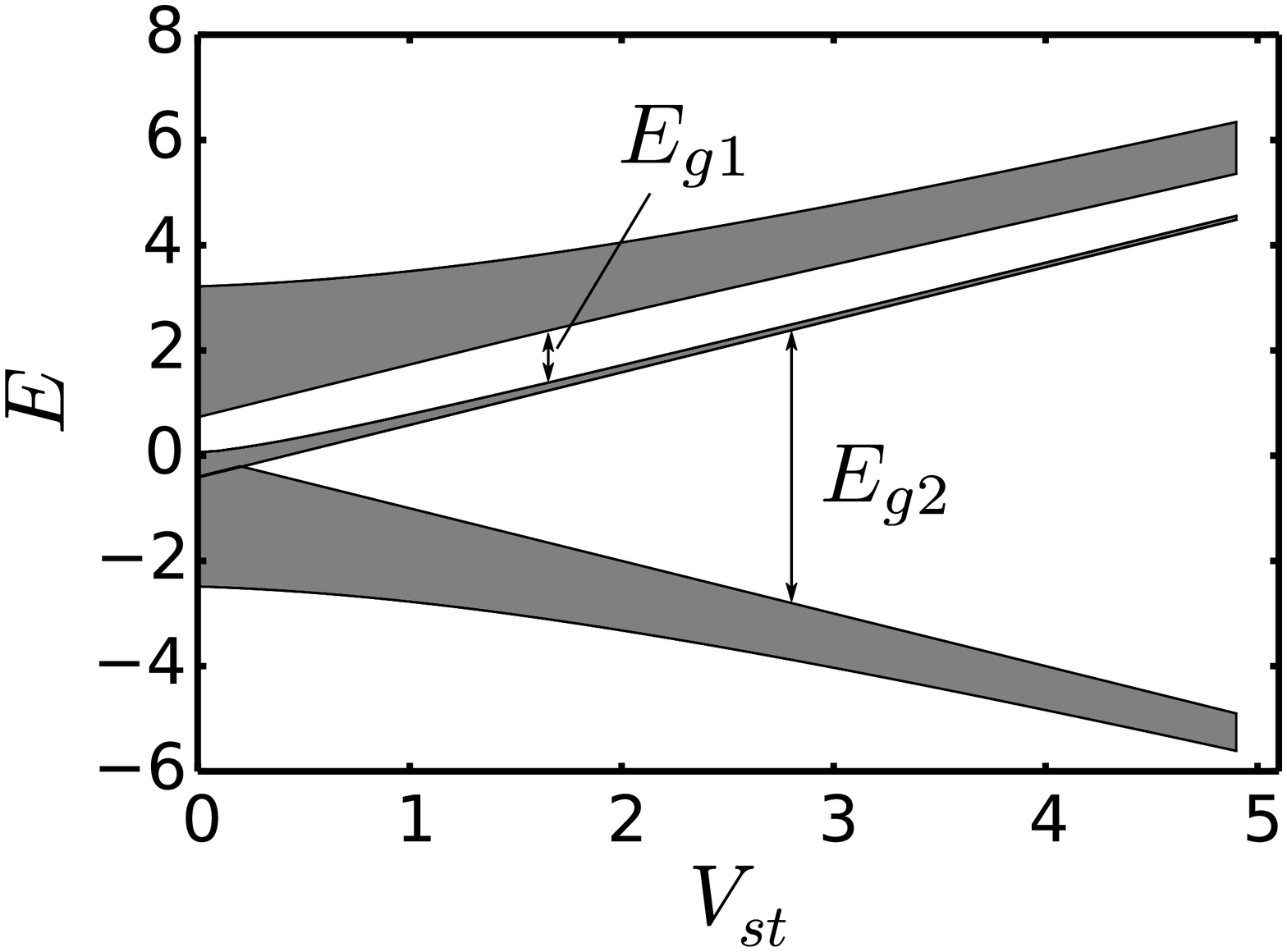}}
\subfloat[][]{\includegraphics[width=1.7in]{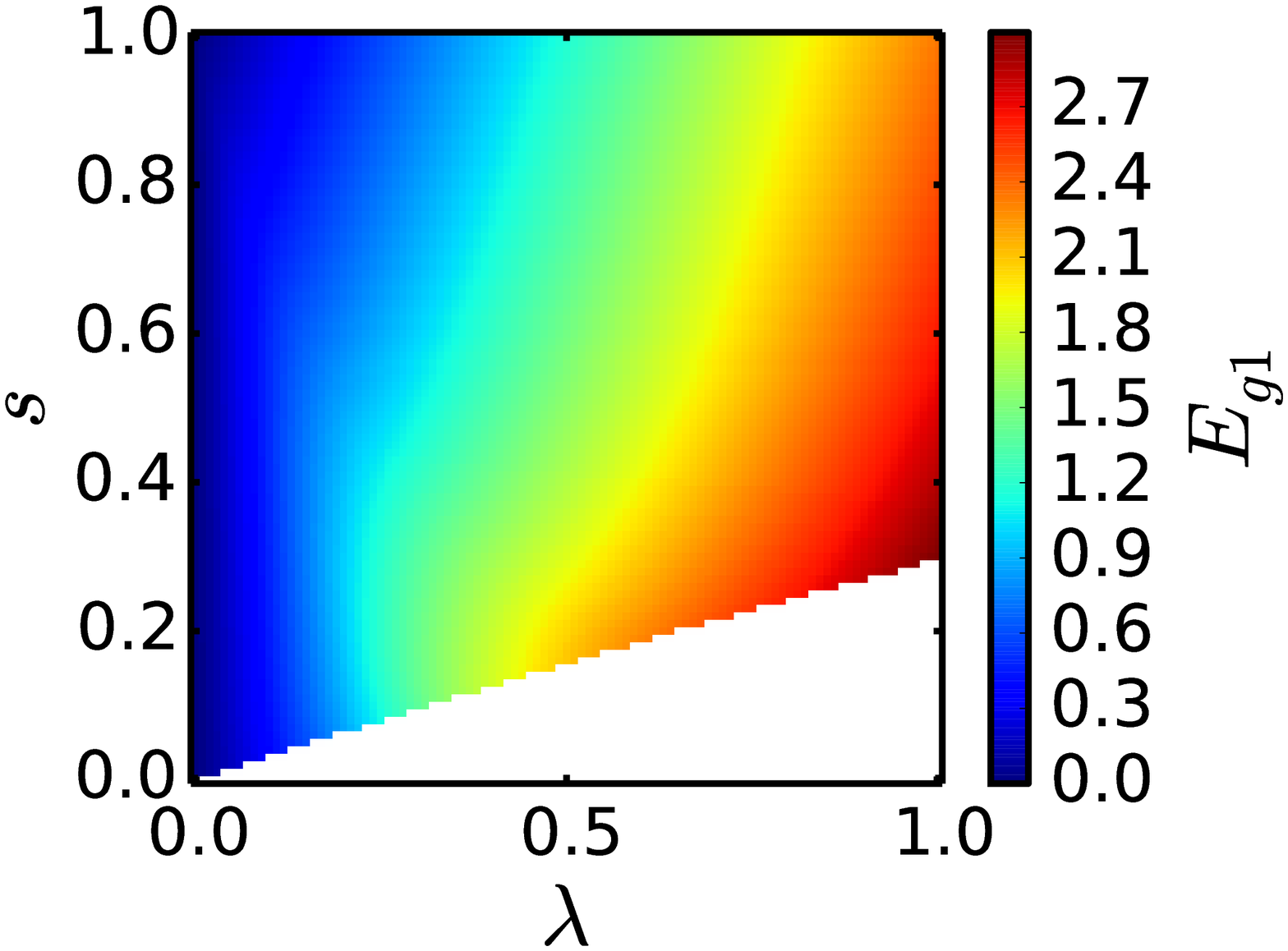}}
\end{minipage}
\caption{(a), (c) Evolution of band structure of Lieb lattice in function of $V_{st}$ for fixed $\lambda=0.2$. (b),(d) Maps of single-particle energy gap $E_{g1}$ depending on $\lambda$ and staggered potential $V_{st}$, parametrized by $V_{st}=\lambda\tan (s\pi/2)$. The top and bottom rows correspond to $\phi=\pi/2$, and $\phi=\pi/4$, respectively. The white line in (b) shows the border between direct (above the line) and indirect (below the line) gap between the upper and middle bands.}
\label{fig:gaps}
\end{figure}

Two energy gaps are indicated in Fig. \ref{fig:transstr}(b) on the right, $E_{g_1}$ between two topologically nontrivial bands, the upper ($C=1$) and the middle band ($C=-1$), and $E_{g_2}$ between topologically nontrivial middle band ($C=-1$) and topologically trivial lower band ($C=0$). We investigate a magnitude of these gaps as a function of model parameters. In Fig. \ref{fig:gaps}(a), a schematic evolution of the energy bands as a function of a staggered sublattice potential $V_{st}$ for $\lambda=0.2$ and $\phi=\pi/2$ is shown. With an increase of $V_{st}$ increases $E_{g_2}$ separating two topologically nontrivial higher energy bands from the lower band. This corresponds also to decoupling of a sublattice indicated by a red color from a sublattice indicated by a blue color in Fig. \ref{fig:transstr}(a). In a limit of $V_{st}\rightarrow\infty$, two sublattices are completely decoupled and the Lieb lattice transforms into the checkerboard lattice, (blue sites in Fig. \ref{fig:transstr}(b)). At the same time, the energy gap $E_{g_1}$ between two topologically nontrivial bands from the checkerboard lattice decreases monotonically to zero. A map of a magnitude of the energy gap $E_{g_1}$ as a function of the staggered sublattice potential $V_{st}$ and $\lambda$ for $\phi=\pi/2$ is shown in Fig. \ref{fig:gaps}(b). The staggered sublattice potential $V_{st}$ is varied from $V_{st}=0$ to $V_{st}\rightarrow\infty$, which can be performed by introducing a parameter $s$ given by a formula $V_{st}=4\tan (s\pi/2)$, where $s$ changes in a range of values $s=(0,1)$. For an isolated checkerboard lattice corresponding to $s=1$ ($V_{st}\rightarrow\infty$),  $E_{g_1}=0$. An introduction of finite $V_{st}$ opens the energy gap $E_{g_1}$. 

For sufficiently high $V_{st}$ the energy gap is a direct gap in M point of the Brillouin zone, with magnitude $E_{g1}=\sqrt{4t^2+V_{st}^2}-V_{st}$ . Below $V_{st}=\frac{t^2}{2\lambda}-2\lambda $ (white line in fig \ref{fig:gaps}(b)) the bottom of the highest band is located at $\Gamma$ point, therefore $E_{g1}$ is an indirect gap of magnitude $4\lambda$. We note that the bandwidth of the middle band in the topologically nontrivial region is also $4\lambda$, so the flatness ratio of the middle band is $\leq 1$. The energy gaps for phase $\pi/4$ show similar behaviour, although closed-form expression for $E_{g1}$ for high $V_{st}$ cannot be obtained. On the other hand, for $\phi=\pi/2$ and $V_{st}\rightarrow \infty$ the bands touch at the whole boundary of the Brillouin zone (hence the energy of highest band at both $M$ and $K$ points asymptotically approach the top of the middle band), while for $\phi=\pi/4$ the gap is closed only at $M$ point.

If additional hoppings are included for $\phi=\pi/4$ (Figs \ref{fig:gaps}(c), (d)), the top of middle band is not located in any high-symmetry point, therefore $E_{g1}$ can be obtained only numerically. In Fig. \ref{fig:gaps}(c) we show dependence of energy gaps on $V_{st}$ for $\lambda=0.2$. Similarily to the previous case, $E_{g2}$ increases to infinity with increased $V_{st}$. However, contrary to the previous case, in the $V_{st}\rightarrow \infty$ limit $E_{g1}$ remains finite (as was noted in Ref\cite{Sun}, additional hoppings open the gap for checkerboard model). As shown in Fig. \ref{fig:gaps}(d), the value of this gap depends on $\lambda$, which is the only single-particle energy scale in $V_{st}\rightarrow \infty$ limit. We note that $E_{g1}$ for given $\lambda$ has maximum for finite $V_{st}$ (Fig. \ref{fig:gaps}(d)), e.g. at $s\approx 0.3$ for $\lambda=0.2$. Therefore the additional atoms increase the energy gap, which may be beneficial for stability of FCI states.

\section{Many-body results}

\subsection{The transition between Lieb and checkerboard lattices}

The staggered sublattice potential $V_{st}$ controls the energetic distance between sites forming a checkerboard lattice (indicated by blue color in Fig. \ref{fig:transstr}(b)) and extra sites (indicated by red color in Fig. \ref{fig:transstr}(b)) introduced to create the Lieb lattice. Analyzing an existence of a Laughlin-like phase during the transition between two lattices, we look at the magnitude of the energy gap between 3-fold degenerate ground states and the fourth state. We perform calculations on $(4\times 6)$ torus for interaction parameters $V_{NN}=1.5$ and $V_{NNN}=1$. Despite the flatness ratio not exceeding one, flat-band approximation \cite{Neupert, zoology} is applied as a first approximation, to focus only on the effects of interaction, neglecting effects of single-particle dispersion and mixing with other bands. In our calculations, we assume the lower band is completely filled and the middle band is filled in 1/3. We have verified the validity of neglecting of excitations from the lower band checking that they do not significantly affect the many-body energy of the three lowest states. We only noticed some effect of electrons from the lower band close to a single particle topological phase transition, where results should be treated tentatively.

\begin{figure}
\begin{minipage}{0.5\textwidth}
  \includegraphics[width=\textwidth]{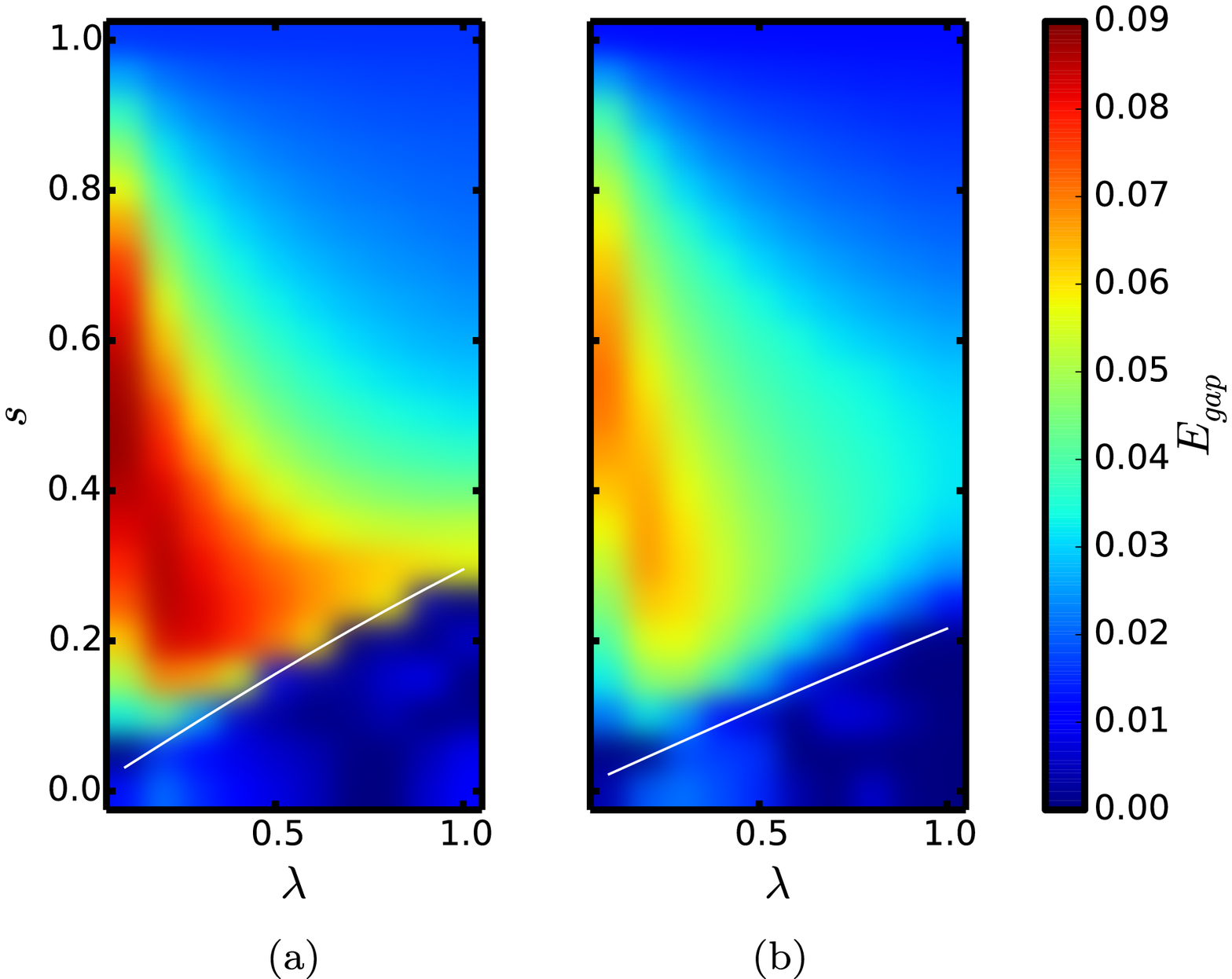}
\end{minipage}
\caption{A map of the energy gap between 3-fold ground state degeneracy and the fourth state for non-flattened middle band with (a) phase $\phi=\pi/2$ and (b) phase $\phi=\pi/4$ (right) as a function of a parameter $\lambda$ and a staggered sublattice potential $V_{st}$ parametrized by $V_{st}=\lambda\tan (s\pi/2)$. Interaction strengths are $V_{NN}=1.5$,$V_{NNN}=1$. The white line denotes the single-particle topological phase transition for $V_{st}=2\lambda$ for $\phi=\pi/2$ and $V_{st}=\sqrt{2}\lambda$ for $\phi=\pi/4$.}
\label{fig:fi2}
\end{figure}

We first consider the situation for $\phi=\pi/2$ in the second term of Hamiltonian given by Eq. \ref{Ham}. Fig. \ref{fig:fi2} on the left shows a map of the energy gap as a function of $\lambda$ and a staggered sublattice potential, represented by the parameter $s$. A single particle topological phase transition is marked by a white line in the graph, with a topologically nontrivial region above the line. Opening of the energy gap coincides with single-particle topological phase transition for $V_{st}=2\lambda$, similarly to results from Ref. \cite{PRX}. Within a topologically nontrivial region the energy spread $\delta$ of 3-fold degenerate ground state does not exceed $\delta=0.015$. Therefore, in a major part of this region 3-fold degenerate ground state separated by the gap is clearly seen in the energy spectrum. Values of the parameter $s\approx 1$ ($V_{st}\rightarrow\infty$) corresponds to an isolated checkerboard lattice giving the energy gap $E_{gap}\approx 0.02$. However, for infinite staggered potential, $s=1$ the energy gap $E_{g1}=0$, and validity of results is uncertain because one cannot restrict calculations to one band only when the gap closes. Also, for $s$ close to 1 the spread of three states becomes comparable with energy gap, therefore their quasi-degeneracy is not visible. For smaller values of a parameter $s$, a region with an increased energy gap appears (a red color area in Fig. \ref{fig:fi2}), with the largest energy gap $E_{gap}\approx 0.08$ for $\lambda\approx 0.1$ and the parameter $s\approx(0.3,0.7)$ ($V_{st}\approx(2.0,8.0)$). Thus, an interaction with extra sites, along with opening a single-particle gap, stabilizes the FCI phase. Interestingly, the maximum values of many-particle gap coincide with the white line in Fig. \ref{fig:gaps}(b) - the transition between indirect and direct gap.

In Fig. \ref{fig:fi2} on the right a phase diagram for a phase $\phi=\pi/4$ is shown. There are no significant qualitative differences comparing to results for $\phi=\pi/2$. Quantitatively, the magnitude of many-particle gap is smaller than for $\phi=\pi/2$. Also, the region of increased gap is slightly bigger than for $\phi=\pi/2$, because the topological phase transition occurs at $V_{st}=\sqrt{2}\lambda$ instead of $V_{st}=2\lambda$.
 
\begin{figure}
\begin{minipage}{0.5\textwidth}
  \includegraphics[width=\textwidth]{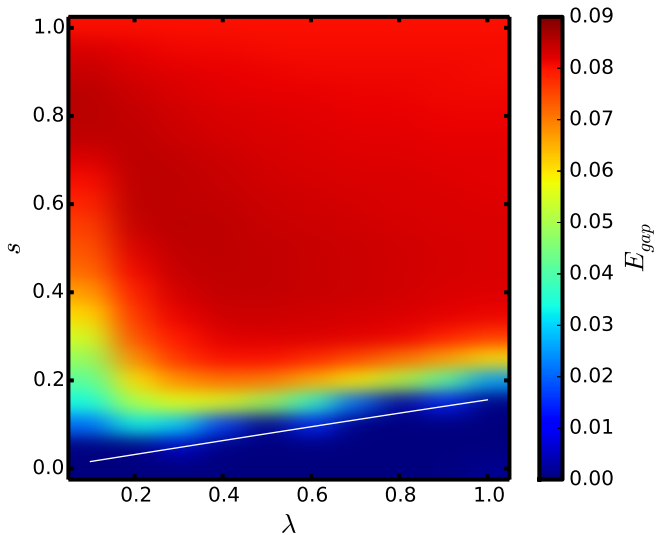}
\end{minipage}
\caption{A map of the energy gap between 3-fold ground state degeneracy and the fourth state for a middle band flattened using additional hoppings $t_2$ and $t_3$, for a phase $\phi=\pi/4$, as a function of a parameter $\lambda$ and a staggered sublattice potential $V_{st}$ parametrized by $V_{st}=\lambda\tan (s\pi/2)$. Interaction strengths are $V_{NN}=1.5$, $V_{NNN}=1$. The white line denotes the single-particle topological phase transition for $V_{st}=\lambda$.}
\label{fig:fi4flat}
\end{figure}

\begin{figure}
\begin{minipage}{0.5\textwidth}
  \includegraphics[width=\textwidth]{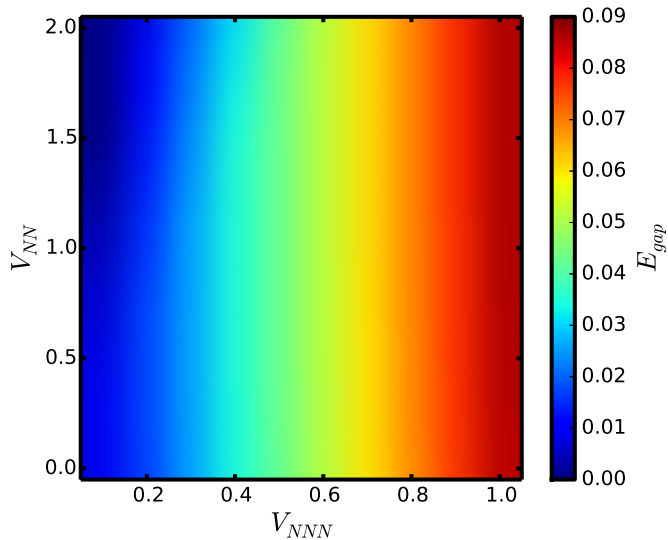}
\end{minipage}
\caption{A map of the energy gap as a function of interaction parameters $V_{NN}$ and $V_{NNN}$ for $\lambda=0.2$, $V_{st}=2$.}
\label{fig:intparams}
\end{figure}

In Fig. \ref{fig:fi4flat}, a phase diagram for $\phi=\pi/4$ with flattened middle band is shown. This corresponds to a map of the single particle energy gap $E_{g1}$ from Fig. \ref{fig:gaps}(d). Within a major part of a parameters range, the energy gap is approximately constant and larger in comparison to the energy gaps for non-flattened bands from Fig. \ref{fig:fi2}, with maximum of $E_{gap}\approx 0.085$. The single-particle gap $E_{g1}$ remains open in the limit $V_{st}\rightarrow\infty$. Finite value of many-particle gap in this limit agrees with earlier results for the checkerboard model \cite{Neupert, PRX}. No gap closing for finite $V_{st}$ shows that FCI on the Lieb lattice with additional hoppings is adiabatically connected to that on checkerboard lattice. A decrease of the energy gap $E_{gap}$ is only seen for $\lambda\approx 0.1$ and close to a single-particle topological phase transition ($V_{st}=\lambda$) marked by a white line.
\\ \indent
In fig ~\ref{fig:intparams} we show the dependence of the energy gap between 3-fold ground state degeneracy and the fourth states on interaction parameters for fixed $\lambda=0.2$ and $V_{st}=2$. In general, the energy gap scales approximately linearly with an interaction between next-nearest neighbors $V_{NNN}$ (an interaction between particles occupying blue color sites in Fig. \ref{fig:mom_sflow}(a)) and only slightly depends on $V_{NN}$ (an interaction between particles occupying sites with different colors in Fig. \ref{fig:mom_sflow}(a)). This is related to the fact that for this choice of parameters the states from the middle band are in $98\%$ localized within the sublattice forming a checkerboard lattice, blue color sites in Fig. \ref{fig:mom_sflow}(a). 

\subsection{Identification of FCI phase}
\begin{figure}
\begin{minipage}{0.5\textwidth}
    \begin{minipage}{0.45\textwidth}
    \subfloat[][]{\includegraphics[width=1.6in]{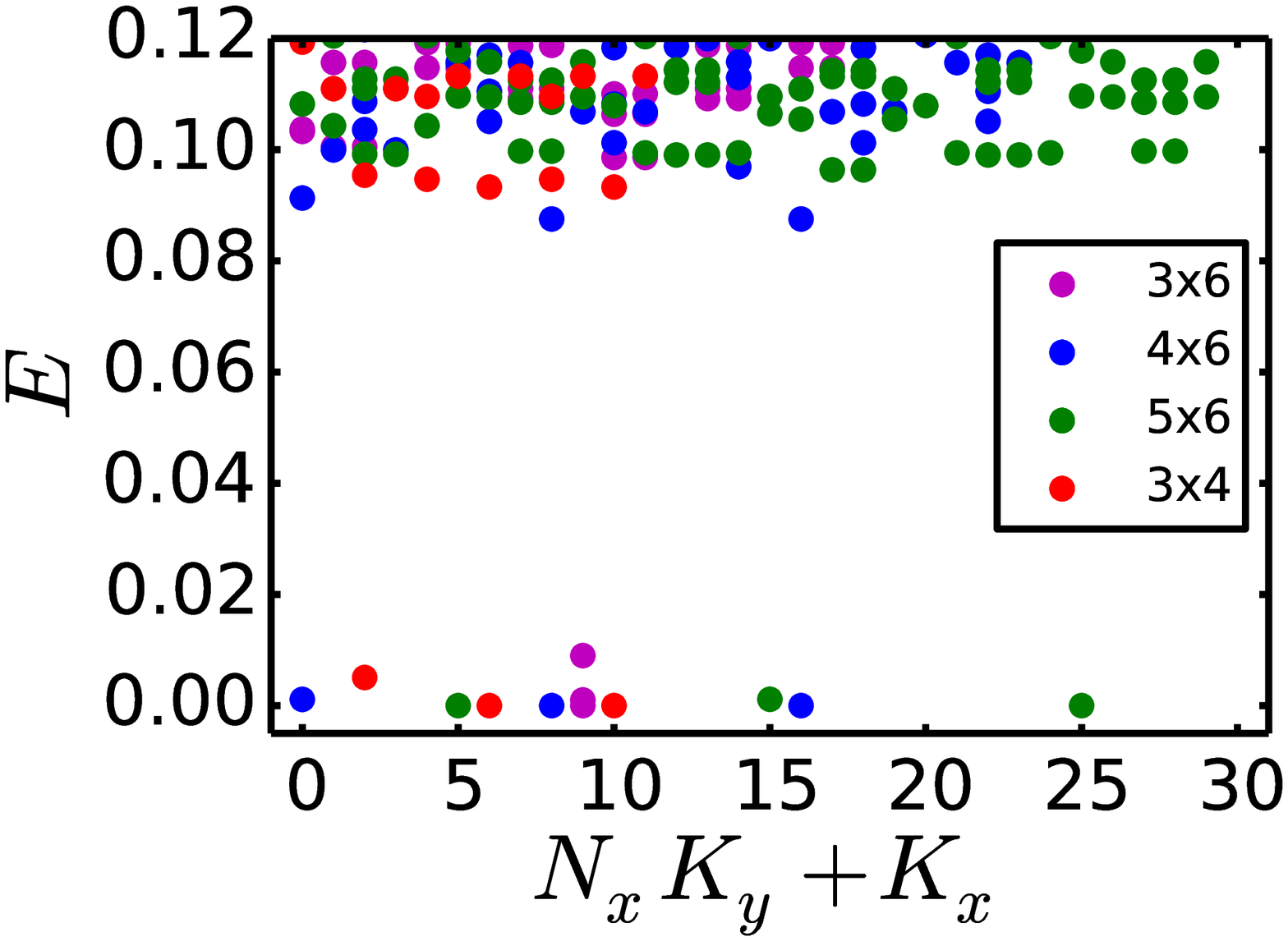}}
    \end{minipage}
    \begin{minipage}{0.45\textwidth}
    \subfloat[][]{\includegraphics[width=1.6in]{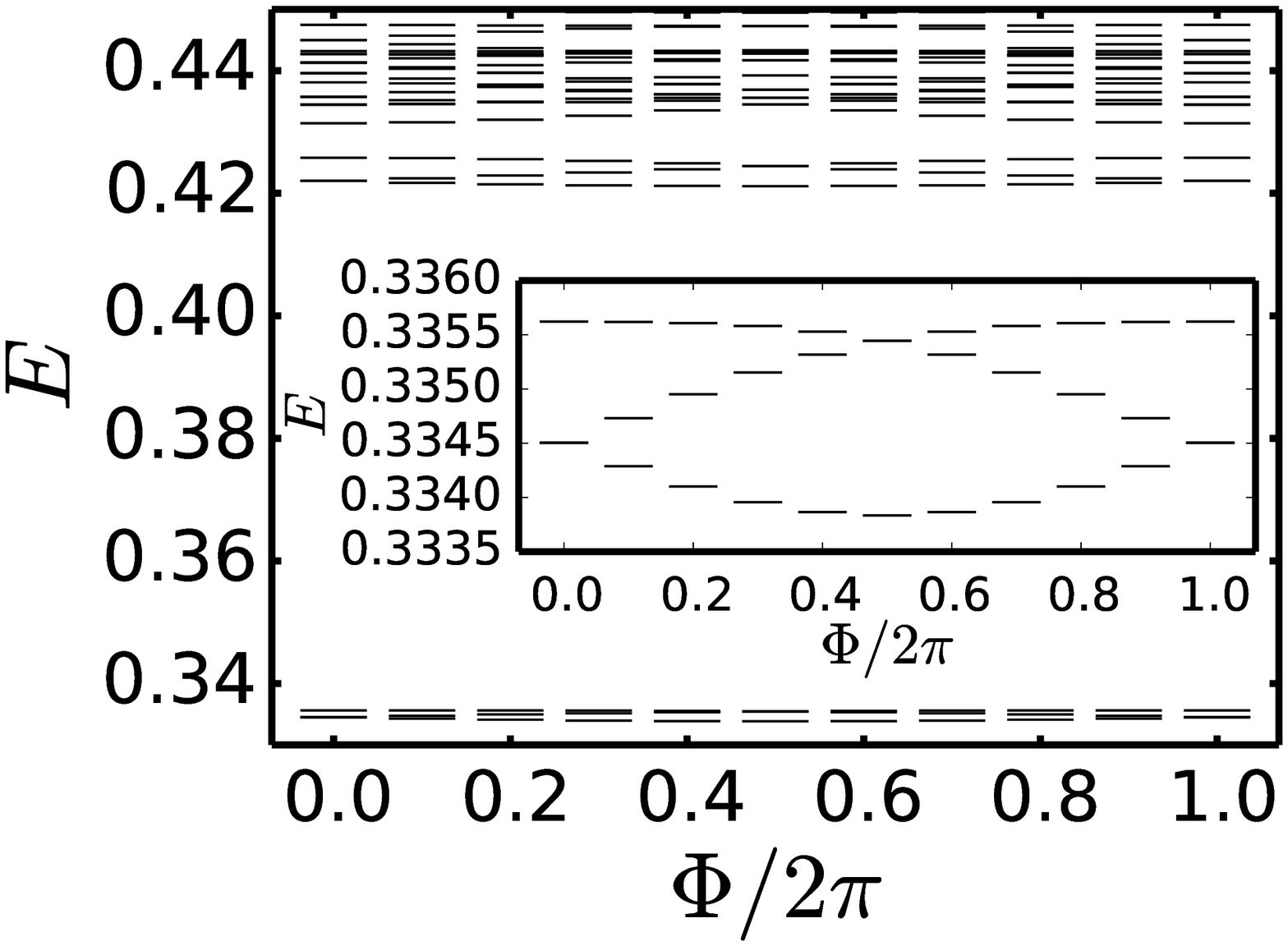}}
    \end{minipage}
\end{minipage}
\caption{a) Momentum-resolved low energy spectra for systems with different sizes given by $(N_x\times N_y)$ for parameters $\lambda=0.2$, $V_{st}=2$, $V_{NN}=1.5$ and $V_{NN}=1$. The energy is rescaled so that ground state energy is set to 0. The momenta of 3 quasi-degenerate states agree with a counting rule for FCI. (b) Spectral flow upon flux insertion for 4x6 lattice. The 3-fold degenerate ground-states flow into each other and do not cross with higher energy states. The inset shows magnified view of the ground state manifold evolution.}
\label{fig:mom_sflow}
\end{figure}

\begin{figure}
\begin{minipage}{0.5\textwidth}
  \includegraphics[width=\textwidth]{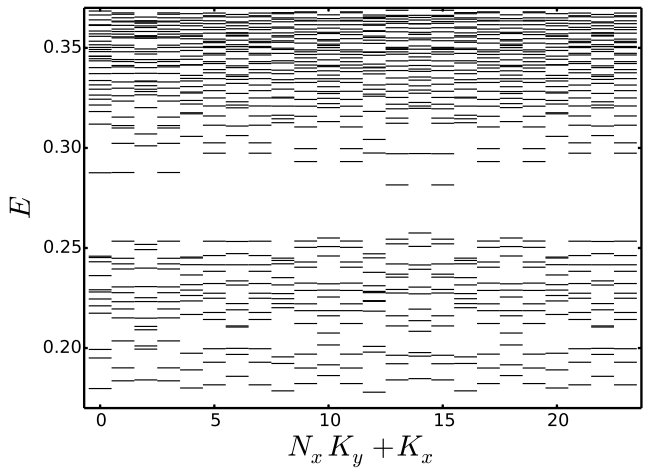}
\end{minipage}
\caption{Momentum-resolved energy spectrum for $N=7$ electrons on $(4\times 6)$ torus for parameters $\lambda=0.2$, $V_{st}=2$,$V_{NN}=1.5$ and $V_{NN}=1$. The number of states below the gap starting around $E=0.25$ is 12 for each momentum sector. This is in agreement with counting for Laughlin quasihole states.}
\label{fig:quasihole}
\end{figure}

We would like to confirm whether red regions with larger energy gaps in Figs. \ref{fig:fi2} and \ref{fig:fi4flat} correspond to FCI phase. Thus, for chosen parameters from this region, $\lambda=0.2$ and $V_{st}=2$, we investigate signatures of $1/3$ Laughlin-like state.  Fig. \ref{fig:mom_sflow}(a) shows momentum-resolved energy spectrum for different torus sizes. The energy spectra are plotted with respect to the ground state energy at $E=0$. We find that for each system size we have 3-fold quasi-degenerate ground state, whose momentum counting corresponds to that obtained from generalized Pauli principle \cite{PRX}. In the case of $N_x\times N_y=(4\times 6)$ this correspond to total momenta of three quasi degenerate ground states for momenta $(K_x,K_y)$: $(0,0)$, $(0,4)$, $(0,8)$. The electron density of the ground state manifold is almost uniformly distributed within sublattice B, as expected for the incompressible liquid. Small variations can be attributed to finite size effects. However, due to localization of single-particle wavefunctions on sublattice B, the sublattice A is significantly less filled. In Fig. \ref{fig:mom_sflow}(b) the spectral flow upon magnetic flux insertion for $(4\times 6)$ torus is shown. The 3-fold degenerate ground states do not intersect with higher states. Three states flow into each other and return to themselves after insertion of three magnetic fluxes. 
This no-mixing property of the ground state manifold with higher energy states is necessary but not sufficient to prove the existence of a Laughlin-like phase. Thus, we analyze quasihole excitations from this state \cite{PRX, zoology}. Figure \ref{fig:quasihole} shows quasihole spectra for $N=7$ electrons on $(4\times 6)$ torus (3 quasiholes). In this case, $12$ quasihole states per momentum sector for Laughlin-like excitations is predicted. This is indeed observed in Fig. \ref{fig:quasihole}. Similarily, the results for 5x5 torus filled by 8 electrons (one quasihole) also obeys the counting rules. The spectrum is divided into two parts separated by a clear energy gap, with $12$ quasihole states per momentum sector below the gap. Thus, our results strongly suggest the presence of FCI in this system.

\section{Conclusions}
In summary, we have analyzed the transition between a checkerboard lattice and a Lieb lattice in the context of FCI phase for $1/3$ filling of a topologically nontrivial energy band. Results were presented for two different complex phases, and a model with a flattened topologically nontrivial band. For the non-flattened bands, the additional sites open the single-particle energy gap and allow FCI to exist. For a flattened band, they increase the single-particle energy gap and stabilize the FCI. The existence of FCI is proven by topological degeneracy, spectral flow and momentum counting, both for exact 1/3 filling and systems with quasiholes. The topologically nontrivial character of FCI phase is also seen by the fact that it exists only in parameter region corresponding to single-particle topologically nontrivial band.

{\it Acknowledgment}. We thank T. Neupert for very constructive remarks and suggestions regarding the identification of the FCI phase. The authors acknowledges partial financial support from the sources granted for science development in the years 2013-2016, Grant No. IP2012 007372. 

\bibliography{article}

\end{document}